\RequirePackage{ifpdf}
\ifpdf 
\documentclass[pdftex]{sigma}
\else
\documentclass{sigma}
\fi

\usepackage{slashed}

\newcommand{\is}{{\Sigma \hspace{-1.1em} \int}}
\begin{document}

\allowdisplaybreaks

\renewcommand{\PaperNumber}{093}

\FirstPageHeading

\renewcommand{\thefootnote}{$\star$}

\ShortArticleName{Heat Trace Asymptotics on Noncommutative Spaces}

\ArticleName{Heat Trace Asymptotics on Noncommutative Spaces\footnote{This paper is a
contribution to the Proceedings of the 2007 Midwest
Geometry Conference in honor of Thomas~P.\ Branson. The full collection is available at
\href{http://www.emis.de/journals/SIGMA/MGC2007.html}{http://www.emis.de/journals/SIGMA/MGC2007.html}}}

\Author{Dmitri V. VASSILEVICH~$^{\dag\ddag}$}

\AuthorNameForHeading{D.V. Vassilevich}

\Address{$^\dag$~Instituto de F\'isica, Universidade de S\~ao Paulo,\\
$\phantom{^\dag}$~Caixa Postal 66318 CEP 05315-970, S\~ao Paulo, S.P., Brazil}
\EmailD{\href{dmitry@dfn.if.usp.br}{dmitry@dfn.if.usp.br}}

\Address{$^\ddag$~V.A.~Fock Institute of Physics, St.~Petersburg University, Russia}

\ArticleDates{Received August 30, 2007; Published online September 25, 2007}

\Abstract{This is a mini-review of the heat kernel expansion for
generalized Laplacians on various noncommutative spaces. Applications to the
spectral action principle, renormalization of noncommutative theories and
anomalies are also considered.}


\Classification{81T75; 58B34}

\rightline{\it Dedicated to the memory of Tom Branson}

\renewcommand{\thefootnote}{\arabic{footnote}}
\setcounter{footnote}{0}

\section{Introduction}
Let us start with a brief description of the standard set-up for
studying the heat trace asymptotics on a commutative space.
Further details can be found in the monograph by Gilkey
\cite{Gbook}. Let $M$ be a compact Riemannian manifold of
dimension $m$, and let $D$ be an operator of Laplace type acting
on the space of smooth sections of a vector bundle $V$ over $M$.
In a local basis any Laplace type operator can be represented as
\begin{gather*}
D=-(g^{\mu\nu}\partial_\mu\partial_\nu +a^\sigma\partial_\sigma +b),
\end{gather*}
where $g^{\mu\nu}$ is the inverse metric on $M$, and $a^\sigma$ and $b$
are matrix-valued functions. There is a~unique connection
$\nabla=\partial+\omega$ on $V$ and
a unique endomorphism $E$ of $V$ such that
\begin{gather}
 D=-(g^{\mu\nu}\nabla_\mu \nabla_\nu +E) .\label{Dcov}
\end{gather}

For a positive $t$ the heat operator $e^{-tD}$ associated to $D$
exists and is trace class. Let $f$ be a smooth smearing function on $M$.
As $t\to +0$ there is a full asymptotic expansion (called the heat kernel
expansion in physics literature)
\begin{gather}
{\rm Tr}(fe^{-tD}) \simeq \sum_{n=0} t^{(n-m)/2} a_n(f,D).
\label{asymptotex}
\end{gather}
The heat kernel coef\/f\/icients $a_n(f,D)$ are {\it locally
computable}, i.e.\ they can be expressed through integrals of local
invariants constructed from the symbol of $D$. If $M$ has no
boundary\footnote{The work by Tom Branson and Peter Gilkey
\cite{Branson:1990xp} devoted to the heat trace asymptotics on
manifolds {\it with} boundary is perhaps the best known to
physicists paper by Tom. This paper was, in a sense, the starting
point of my collaboration with Tom. In \cite{Vassilevich:1994we} I
corrected a minor error of \cite{Branson:1990xp} made in the $a_4$
coef\/f\/icient for mixed boundary conditions. The reaction of Tom and
Peter was very friendly, and they invited me to join them and
calculate the next coef\/f\/icient $a_5$. We performed this
calculation \cite{Branson:1995cm}, though with some restrictions
on the boundary conditions (pure Dirichlet or Neumann), and on the
manifold (domain in f\/lat space or totally geodesic boundaries).
The restrictions on the manifold were removed by Klaus Kirsten
\cite{Klausa5}. Later, when Klaus joined the collaboration, it
became possible to do $a_5$ for generic mixed boundary conditions
\cite{Branson:1999jz}.}, $\partial M=\varnothing$, the odd-numbered
coef\/f\/icients vanish, $a_{2k-1}(f,D)=0$, while leading
even-numbered coef\/f\/icients read
\begin{gather}
a_0(f,D)=(4\pi)^{-m/2} \int_{M}d^mx\, \sqrt{g}\, {\rm tr}_V
(f),\label{a0}\\
a_2(f,D)=(4\pi)^{-m/2}\frac 16 \int_{M}d^mx\, \sqrt{g}\, {\rm tr}_V
\left( f(6E+R)\right),\label{a2}\\
a_4(f,D)=(4\pi)^{-m/2} \frac 1{360} \int_{M}d^mx\, \sqrt{g}\, {\rm tr}_V
\left(f(60\nabla^2E+60RE+180E^2 \right.\nonumber\\
\left. \phantom{a_4(f,D)=}{} +12\nabla^2R +5R^2 -2R_{\mu\nu}R^{\mu\nu}
+2R_{\mu\nu\rho\sigma}R^{\mu\nu\rho\sigma} + 30\Omega_{\mu\nu}
\Omega^{\mu\nu} )\right) .\label{a4}
\end{gather}
Here $\Omega_{\mu\nu}=\partial_\mu\omega_\nu -\partial_\nu\omega_\nu
+\omega_\mu\omega_\nu -\omega_\nu\omega_\nu$ is the bundle curvature,
$R_{\mu\nu\rho\sigma}$, $R_{\mu\nu}$ and $R$ are the Riemann curvature
tensor, the Ricci tensor and the curvature scalar, respectively.  Later
on we shall use these formulae to compare them to the noncommutative (NC)
case.

The coef\/f\/icients $a_n$ carry a lot of important mathematical
information about the bundle $V$ and the operator $D$. The Index
is perhaps the most notable example. The signif\/icance of these
coef\/f\/icients to physics is even greater. The coef\/f\/icients $a_n$
def\/ine the one-loop divergences in quantum f\/ield theory, the
quantum anomalies, some expansions of the ef\/fective action, etc.
For more information the interested reader may consult
\cite{Gold,Gbook,Kbook,Vassilevich:2003xt,Vassilevich:2004id}.

The aim of this paper is to show how and to which extent this
scheme can be extended to noncommutative spaces. To describe a
noncommutative deformation of a given manifold $M$ one takes the
algebra $\mathcal{A}$ of smooth functions on $M$ and deforms it to
an algebra $\mathcal{A}_\Theta$, which will be assumed associative
but not commutative. Usually, such a deformation can be realized
again on smooth functions on $M$, but with the point-wise product
replaced by a noncommutative product.

This paper is organized as follows. In the next section we brief\/ly
discuss the spectral action principle which is one of the most
important application of heat kernel expansion on NC mani\-folds.
Then we turn to the Moyal type NC products and describe
corresponding heat trace asymptotics in some detail. Other NC
spaces are considered in Section~\ref{sec-other}. Section~\ref{sec-ren} contains a couple of examples of the use of the heat
kernel in renormalization and calculations of the anomalies.
Concluding remarks are given in the last section.

\section{Application: spectral action principle}\label{sec-sac}
In noncommutative geometry \cite{ConnesBook} there is one specif\/ic
application of the heat kernel expansion which is related to the
spectral action principle \cite{Chamseddine:1996zu}. This action
is constructed from the spectral triple
$(\mathcal{A},\mathcal{H},\slashed{D})$ consisting of an algebra
$\mathcal{A}$ acting on a Hilbert space $\mathcal{H}$ and
of a Dirac operator~$\slashed{D}$ which describes f\/luctuating f\/ields
(e.g., $\slashed{D}$ may depend on a gauge connection). The spectral
action for these f\/luctuating f\/ields is then def\/ined as
${\rm Tr}(\Phi (\slashed{D}^2/\Lambda^2))$, where $\Phi$ is a
positive even function, $\Lambda$ is a mass scale parameter.
 For large $\Lambda$ the spectral action can be expanded as
\begin{gather*}
{\rm Tr}(\Phi (\slashed{D}^2/\Lambda^2)) \simeq
\sum_{k=0} \Lambda^{m-k} \Phi_{k} a_k(1,\slashed{D}^2),
\end{gather*}
where $\Phi_k$ are def\/ined through the Laplace transform of $\Phi$
(see \cite{Nest:2002ba}). The Dirac operator squared is a Laplace
type operator. Therefore, the heat kernel expansion allows to
evaluate the $\Lambda$ expansion of the spectral action. Note,
that direct calculations of the spectral action without use of the
heat kernel are very cumbersome (though possible, cf.~\cite{Essouabri:2007wi}). Some examples of spectral actions will
be brief\/ly discussed below.
\section{Moyal spaces}
The most popular example of a noncommutative associative product
is the Moyal product (also known as the Groenewold--Moyal or
Moyal--Weyl product) which can be written as
\begin{gather}
(f\star g)(x)=\exp \left( \frac i2 \Theta^{\mu\nu} \partial_\mu^x
\partial_\nu^y \right) f(x)g(y)\vert_{y=x}, \label{Moyal}
\end{gather}
where $\Theta$ is a constant skew-symmetric matrix. As it stands,
the product (\ref{Moyal}) is valid for smooth functions, but by
going to the Fourier transforms one can extend this def\/inition to
the case when at least one of the functions is less smooth. The
def\/inition (\ref{Moyal}) assumes existence of a~global coordinate
system on $M$ at least in the NC directions. The properties of the
heat kernel expansion depend crucially on the compactness (or
non-compactness) of these NC directions.

From the def\/inition (\ref{Moyal}) it follows that
\begin{gather*}
e^{ikx}\star f(x)=f(x+\Theta k/2)e^{ikx},\qquad f(x)\star e^{ikx}=
f(x-\Theta k/2)e^{ikx}.
\end{gather*}

A suitable generalization of the Laplace operator (\ref{Dcov}) for
Moyal spaces which covers most of the applications reads
\begin{gather}
D=-(g^{\mu\nu}\nabla_\mu\nabla_\nu + E), \qquad \nabla_\mu =
\partial_\mu +\omega_\mu ,\label{op1}\\
\omega_\mu = L(\lambda_\mu)-R(\rho_\mu),\label{op2}\\
E=L(l_1)+R(r_1)+L(l_2)R(r_2),\label{op3}
\end{gather}
where $L$ and $R$ are left and right Moyal multiplications,
$f_1\star f_2= L(f_1)f_2=R(f_2)f_1$. External f\/ields
$l_{1,2}$, $r_{1,2}$, $\lambda_\mu$ and $\rho_\mu$ are supposed to
be smooth and rapidly decaying at the inf\/inity (if $M$ has
non-compact dimensions). In this section (except for one Subsection~\ref{sec-iso}) we assume that the metric
$g^{\mu\nu}$ is constant. One can add, if needed, more terms
consisting of the multiplication operators to the right hand sides
of (\ref{op2}) and (\ref{op3}).

The smearing function in (\ref{asymptotex}) should also be
replaced by a product of Moyal multiplications, so that we shall
consider asymptotic properties of the expression
\begin{gather*}
K(l,r,D)={\rm Tr}(L(l)R(r)e^{-tD}).
\end{gather*}
Our general strategy of calculation of the heat kernel expansion
is to isolate the free Laplacian $\Delta=
-g^{\mu\nu}\partial_\mu\partial_\nu$ in $D$, keep $e^{-t\Delta}$
and expand the rest of the exponential. When applying this
procedure, one has to evaluate the asymptotic expansion of the
expressions like
\begin{gather}
T(\bar l,\bar r):={\rm Tr}(L(\bar l)R(\bar
r)e^{-t\Delta}),\label{Tlr}
\end{gather}
where $\bar l$ and $\bar r$ are some polynomials of the smearing
functions, background connections $\lambda$ and~$\rho$, the
potential $E$ and their derivatives. Note, that the corresponding
expressions with only left or only right Moyal multiplications, $
{\rm Tr}(R(\bar r)e^{-t\Delta})$ and  ${\rm Tr}(L(\bar
l)e^{-t\Delta})$ do not depend on $\Theta$, and, therefore, the
heat trace asymptotics for the operators with just one type of the
multiplications look as in the commutative case (see (\ref{a0})--(\ref{a4}) above) up to replacing ordinary products by star
products \cite{Vassilevich:2003yz,GI:2004}.

In generic case, i.e.\ when both $\bar l$ and $\bar r$ are present,
in order to evaluate the trace in (\ref{Tlr}) one has to sandwich the
expression under the trace between two plane waves $e^{ikx}$ and
integrate (or sum up, in the compact case) over the momenta $k$.
It is convenient to make the Fourier transform of $\bar l$ and
$\bar r$. After performing some obvious integrations one gets (see
\cite{Vassilevich:2005vk} for details)
\begin{gather}
T(\bar l,\bar r)\sim \is d^mk \is d^mq \, e^{-tk^2} \bar l_{-q}
\bar r_q e^{-ik\wedge q} ,\label{T2}
\end{gather}
where the symbol ${{\Sigma} \hspace{-0.8em} \int}$ denotes an
integral (resp., a sum) over the momenta for noncompact (resp.,
for compact) dimensions. $\bar l_q$ and $\bar r_q$ denote Fourier
components of $\bar l$ and $\bar r$, and $k\wedge q:=k_\mu
\Theta^{\mu\nu} k_\nu$. We dropped an irrelevant overall factor in
(\ref{T2}). Now we are ready to discuss the small $t$ asymptotic
expansion of $T(\bar l,\bar r)$. The result depends crucially on
compactness of the manifold~$M$.

\subsection{Moyal plane}

Let us consider the Moyal plane $\mathbb{R}^m_\Theta$. The
integral over $k$ in (\ref{T2}) can be easily performed yielding~\cite{Vassilevich:2005vk}
\begin{gather}
T(\bar l,\bar r)=(4\pi t)^{-m/2} \int d^mq\ \bar l_{-q} \bar r_{q}
\exp \left( -\frac 1{4t} \Theta^{\mu\rho}\Theta^\nu_{\ \rho}q_\mu
q_\nu \right) .\label{T3}
\end{gather}
If $\Theta\equiv 0$ one can return in the equation above to the
coordinate space. The resulting expression is just a smeared
heat kernel for the free Laplacian on the commutative plane.
Therefore, we simply conf\/irm a correct value for $a_0(lr,\Delta)$,
see (\ref{a0}).

Let us assume that $\Theta$ is non-degenerate (which implies that
$m$ is even). Then the exponent in (\ref{T3}) provides a suppression
of contributions from large $q$. Therefore, to evaluate the behavior
of (\ref{T3}) at small $t$ one has to expand $\bar l_{-q} \bar r_{q}$
in Taylor series around $q=0$ and then integrate over~$q$. One
obtains the following expression \cite{Vassilevich:2005vk}
\begin{gather}
T(\bar l,\bar r)=(\det \Theta)^{-1} (\bar l_0 \bar r_0 +
\mathcal{O} (t)).\label{T4}
\end{gather}
This expression is nothing else than the heat kernel expansion
for the free Laplacian $\Delta$. Comparing (\ref{T4}) to the
the expansion (\ref{asymptotex}) in the commutative case, one
observes some important dif\/ferences. The expansion (\ref{T4})
(i) does not contain negative powers of $t$, (ii) is non-local,
and (iii) is divergent in the limit $\Theta\to 0$.

For applications it is important to know the asymptotic expansion
on the unsmeared heat trace ${\rm Tr}(e^{-tD})$. Such an expansion
cannot be obtained by simply taking $l=r=1$ in the expressions above
since constant functions on $\mathbb{R}^m$ are not Schwartz class.
Moreover, ${\rm Tr}(e^{-tD})$ contains a trivial volume divergence
which must be regularized away by subtracting the (unsmeared) heat
trace of $\Delta$. The methods described above are still applicable,
and after some calculations one obtains  \cite{Vassilevich:2005vk}
\begin{gather*}
{\rm Tr} (e^{-tD}-e^{-t\Delta}) \simeq \sum_{n=2} t^{(n-m)/2}
a_n^{\rm sub}(D),\\
a_n^{\rm sub}(D)=a_n^L(D)+a_n^R(D)+a_n^{\rm mix}(D),\nonumber
\end{gather*}
where the coef\/f\/icients $a_n^{L}$ (respectively, $a_n^R$) depend on left
(respectively, on right) f\/ields
only. In terms of the f\/ields appearing in (\ref{op2}), (\ref{op3})
two leading coef\/f\/icients read
\begin{gather}
a_2^L(D)=(4\pi)^{-m/2} \int d^mx\, l_1(x),\label{a2RmL}\\
a_4^L(D)=(4\pi)^{-m/2}\frac 1{12}
\int d^mx\, (6l_1\star l_1 + \Omega_{\mu\nu}^L\star
\Omega^{L\mu\nu}),\label{a4RmL}
\end{gather}
where $\Omega_{\mu\nu}^L=\partial_\mu\lambda_\nu - \partial_\nu
\lambda_\mu +\lambda_\mu \star \lambda_\nu - \lambda_\nu \star
\lambda_\nu$. The coef\/f\/icients $a_2^R$ and $a_4^R$ are obtained
from (\ref{a2RmL}) and (\ref{a4RmL}) by replacing
$(l_1,\lambda,\Omega^L)$ with $(r_1,\rho,\Omega^R)$. (Note, that
$a_0^{L,R}=0$ due to the subtraction of the free heat kernel). The
coef\/f\/icients $a^L$ and $a^R$ are called planar because of their
similarity to planar diagrams of quantum f\/ield theory.

Mixed coef\/f\/icients $a_n^{\rm mix}$ vanish for $n\le m$. The f\/irst
non-zero coef\/f\/icient is
\begin{gather*}
a_{m+2}^{\rm mix}(D)=\frac {(\det \Theta)^{-1}}{(2\pi)^m} \left( \int
d^mx l_2(x) \int d^my r_2(y) +2 \int d^mx \lambda_\mu (x) \int
d^my \rho^\mu (y)\right). 
\end{gather*}
This coef\/f\/icient is non-local and divergent in the limit $\Theta\to 0$,
as expected.

If $\Theta$ is degenerate, there can be mixed non-local heat
kernel coef\/f\/icients $a_n^{\rm mix}$ for $n\le m$. For the reasons
explained below, this can lead to problems with the
renormalizability  \cite{Gayral:2004cu}.

Heat kernel expansion for non-minimal operators, such as appear in
NC gauge theories, was constructed in \cite{Strel:2006}.
\subsection{NC torus}
Let us now turn to the noncommutative torus $\mathbb{T}_\Theta^m$.
This space was f\/irst constructed by Con\-nes~\cite{ConnesTm}. The
heat kernel expansion on $\mathbb{T}^m_{\Theta}$ was analyzed in
\cite{Gayral:2006vd}. Below we give a brief overview of the
results. On $\mathbb{T}^m_{\Theta}$ it is not that essential
whether $\Theta$ is degenerate or not, but instead some number
theory aspects play an important role.

One can bring $T(\bar l,\bar r)$ to the form
\begin{gather}
T(\bar l,\bar r)=\sqrt{g} (4\pi t)^{-m/2}\sum_{q\in \mathbb{Z}^m}
\sum_{k\in \mathbb{Z}^m}\bar l_{-q} \bar r_q
e^{-\frac {|\theta q -2\pi k|^2}{4t}} ,\label{TT1}
\end{gather}
where $g$ is determinant of a (constant) metric on $\mathbb{T}^m$.
Each term under the sum in (\ref{TT1}) is exponentially small at
$t\to 0$ unless $\theta q -2\pi k=0$ for some $q$ and $k$. However,
this ``smallness'' can be overcompensated by an inf\/inite
number of terms in the sum. To understand better what is going
on, let us consider the case when $\Theta$ has a block-diagonal
form
\begin{gather}
\Theta = \bigoplus_{\theta_i \in 2\pi \mathbb{Q}} \theta_i
\left( \begin{array}{cc} 0 & 1 \\ -1 & 0 \end{array}\right)
\bigoplus_{\theta_j \in \mathbb{R}\backslash 2\pi \mathbb{Q}}
\theta_j \left( \begin{array}{cc} 0 & 1 \\ -1 & 0 \end{array}\right),
\label{Theta}
\end{gather}
where we separated rational and irrational (relative to $2\pi$)
values of the NC parameter. To avoid the overcompensations
mentioned above  we have to
assume that irrational $\theta_j$ satisfy the Diophantine
condition, i.e.\ that there are two constants $C>0$ and $\beta \ge
0$ such that for all nonzero $q\in \mathbb{Z}$
\begin{gather*}
\inf_{k\in \mathbb{Z}} |\theta_j q-2\pi k|\ge \frac
C{|q|^{1+\beta}}. 
\end{gather*}
Generic form of the Diophantine condition which does not rely on a
particular block-diagonal form (\ref{Theta}) of $\Theta$ can be
found in \cite{Gayral:2006vd}. The Diophantine condition means that
$\theta_j$ cannot be too well approximated by rational numbers and
controls behavior of the exponent in (\ref{TT1}).
If this condition is violated, the
asymptotics of the heat trace are unstable (see Appendix B of
\cite{Gayral:2006vd}), and, therefore, this case is not considered
here.

Next we def\/ine a trace (which is proportional to the Dixmier trace)
\begin{gather*}
{\rm Sp} (L(l)R(r)) = \sqrt{g} \sum_{q\in \mathcal{Z}}
l_{-q}r_q, 
\end{gather*}
where the set $\mathcal{Z}\subset \mathbb{Z}^m$ consists of all
$q$ such that $(2\pi)^{-1} \Theta q \in \mathbb{Z}^m$. Then, one
can demonstrate, that
\begin{gather*}
T(\bar l,\bar r)={\rm Sp}(L(\bar l)R(\bar r))
\end{gather*}
up to exponentially small terms in $t$. Moreover, in terms of this trace
the heat kernel coef\/f\/icients look precisely as in the commutative
case! In particular, all odd-numbered coef\/f\/icients vanish, and
\begin{gather}
a_0(l,r,D)=(4\pi)^{-m/2} {\rm Sp}(L(l)R(r)),\nonumber\\
a_2(l,r,D)=(4\pi)^{-m/2} {\rm Sp}(L(l)R(r)E),\label{hkTm}\\
a_4(l,r,D)=(4\pi)^{-m/2}\frac 1{12} {\rm Sp}(L(l)R(r)
(6E^2+2\nabla^2E + \hat \Omega^{\mu\nu}\hat \Omega_{\mu\nu})).
\nonumber
\end{gather}
Here $\hat \Omega_{\mu\nu}=\nabla_\mu\nabla_\nu
-\nabla_\nu\nabla_\mu =
L(\Omega_{\mu\nu}^L)-R(\Omega_{\mu\nu}^R)$.

Let us stress that {\it all} heat kernel coef\/f\/icients are
non-local.

By using the results described above one can calculate the
spectral action for a $U(1)$ gauge f\/ield on $\mathbb{T}^4_\Theta$
\cite{Gayral:2006vd}. Rather unexpectedly, all non-local terms
disappear, and one gets the standard action of NC QED.

\subsection{Finite-temperature manifolds}
To describe f\/inite temperature ef\/fects in quantum f\/ield theory one
often uses the imaginary time formalism where the Euclidean time
variable is periodic with the period equal to the inverse
temperature. For this reason, one studies NC quantum f\/ield theory
on $\mathbb{R}^3\times S^1$. The most interesting case is, of
course, when a compact coordinate on $S^1$ does not commute with a~non-compact coordinate. The heat kernel expansion for the operator
(\ref{op1}) with $\rho=\lambda=0$ was considered in
\cite{Strelchenko:2007xh} assuming for simplicity that non-compact
coordinates on $\mathbb{R}^3$ commute between themselves. It was
found that planar coef\/f\/icients still have the same form as above,
see~(\ref{a2RmL}) and~(\ref{a4RmL}). In the mixed sector
odd-numbered non-local coef\/f\/icients appear. Even-numbered mixed
coef\/f\/icients vanish in this case.

\subsection{Isospectral deformation manifolds}\label{sec-iso}
Isospectral deformation \cite{CL,CDB} is a way to def\/ine a
Moyal-like star product on curved manifolds. Suppose, we have
several commuting isometries acting on a manifold. Then one can
def\/ine a~star product which includes derivatives only with respect
to the coordinates which correspond to the isometries. One can do
some spectral theory on such manifolds as well. In particular, one
can analyze the Dixmier traces \cite{Gayral:2005ad}, analyze the
mixing between ultra violet and infra red scales by the heat
kernel methods \cite{Gayral:2004cs}, and calculate some of the
heat kernel coef\/f\/icients \cite{Gayral:2006vd}. It is interesting
to note, that if the action of the isometries has f\/ixed points,
there are specif\/ic contributions to the heat kernel expansion
coming from these points or submanifolds.

One can def\/ine a Moyal-like star product also for a position
dependent noncommutativity, but then $\Theta$ must be degenerate
\cite{Gayral:2005ih}. The heat kernel expansion was analyzed on
such manifolds~\cite{Gayral:2005ih} by using the methods of
covariant perturbation theory \cite{Barvinsky:1990up}.

\section{Other NC spaces}\label{sec-other}

\subsection{Noncommutativity and the standard model of elementary
particles} Noncommutative geometry can provide an explanation of
the spectrum of particles in the standard model. In the works
devoted to this approach it is assumed that space-time is the
product of a four-dimensional manifold by a f\/inite noncommutative
space. The heat-kernel part of the calculations of the spectral
action is quite standard. Therefore, we do not go into details
here. It is worth noting that the coincidence with the observed
properties of elementary particles is remarkable. The interested
reader may consult the paper \cite{Chamseddine:2007ia} and
references therein.

\subsection{Field theories with an oscillator potential}\label{sec-op}
To achieve renormalizability of $\phi^4$ theory on $\mathbb{R}^4_\Theta$
at all orders of perturbation theory one has to modify the
action by adding an oscillator potential term \cite{Grosse:2004yu}
\begin{gather*}
S_{\rm o.p.}=\frac {\Omega^2}2 \int d^4x (\tilde x^\mu \phi)\star
(\tilde x_\mu \phi),
\end{gather*}
where $\tilde x_\mu =2(\Theta)^{-1}_{\mu\nu} x^\nu$, and $\Omega$
is a parameter. This term leads to an additional potential in~$D$
which behaves as $x^2$. In the presence of a potential which grows
at inf\/inity the heat kernel expansion is modif\/ied essentially even
in the commutative case. In the NC case, the heat kernel expansion
was constructed in \cite{Grosse:2006hh,Grosse:2007dm} and used to
study an induced gauge f\/ield action (in the spirit of the spectral
action principle, see Section~\ref{sec-sac}).

\subsection{Fuzzy spaces}
Roughly speaking, fuzzy spaces are constructed by taking the
harmonic expansion on a usual compact space and truncating it at a
certain level. This truncated set of the harmonics is not closed
under the usual point-wise product, and the most non-trivial part
of this construction is to introduce a new associative product on
this set which tends to the usual product when the truncation goes
to inf\/inity. For example, in the case of fuzzy $S^2$
\cite{Madore:1991bw} this is done in the following way. Let
$\tilde x^a$ be coordinates on $\mathbb{R}^3$ subject to the
restriction $\sum_{a=1}^3(\tilde x^a)^2=1$ (so that $\tilde x^a$
become coordinates on $S^2$). Any function on $S^2$ can be
expanded in a sum of homogeneous polynomials of $\tilde x^a$. By
restricting the order of these polynomials to less than some
number $n$ one obtains functions on a fuzzy version of $S^2$. A
suitable multiplication is obtained by identifying $\tilde x^a$
with certain elements of the algebra of complex $n\times n$
matrices. The restriction to lowest harmonics makes it hard to
distinguish points on the two-sphere, so it becomes ``fuzzy''.

From the point of view of the heat kernel expansion one property of the
fuzzy spaces is most important: the spaces of functions are f\/inite dimensional.
Consequently, the heat trace ${\rm Tr}(e^{-tD})$ becomes a f\/inite sum,
and in the limit $t\to 0$ no inverse powers of $t$ may appear. However,
Sasakura \cite{Sasakura:2004dq,Sasakura:2005px} has demonstrated numerically
that the expansion (\ref{asymptotex}) appears for intermediate values of
$t$. Correct heat kernel coef\/f\/icients (\ref{a0})--(\ref{a4}) were reproduced
both for $f=1$~\cite{Sasakura:2004dq} and $f\ne {\rm const}$~\cite{Sasakura:2005px}
for a number of fuzzy spaces. Analytical understanding of this fact
is still missing.

\subsection{Curved Moyal spaces}\label{sec-curM}
The most obvious way to model a curved Moyal space is to replace
in (\ref{op1}) the constant metric~$g^{\mu\nu}$ by an operator
$L(g^{\mu\nu})$ with a metric $g^{\mu\nu}(x)$ which depends on NC
coordinates. Spectral properties of such operators in a
two-dimensional case were analyzed in \cite{Vassilevich:2004ym}.
Though one can def\/ine suitable Dirac and Laplace operators and
formulate relations between them (e.g., a gene\-ralization of the
Lichnerowicz formula), practical calculations of the heat kernel
coef\/f\/icients become very cumbersome. The best one can do is to
impose the conformal gauge condition on the metric,
$g^{\mu\nu}=e^{-2\rho}\delta^{\mu\nu}$, and calculate the
coef\/f\/icient $a_2$ as power series expansion in the conformal
factor $\rho (x)$. There seems to be no obvious relation of this
coef\/f\/icient to geometric invariants. It could be that dif\/f\/iculties
with analyzing spectral properties of operators on curved Moyal
spaces are related to known problems with formulation of a gravity
theory on such spaces (for a review, see \cite{Szabo:2006wx}).

From the technical point of view, the dif\/f\/iculties with
constructing the heat kernel expansion in the case under
consideration are caused by deformation of the leading symbol (the
part with the highest explicit derivatives) of the Laplacian.
Similar problems appear in another version of NC gravity which
considers $g^{\mu\nu}$ which as a matrix in some internal space
for each pair $(\mu,\nu)$ (see, \cite{Avramidi:2005rn}). Relevant
operators become non-minimal (strictly speaking, they are not
Laplacians any more). Despite some recent progress
\cite{Avramidi:2001tx}, actual calculations of the heat trace
asymptotics for such operators remain a complicated task.

\section{Application: renormalization and anomalies}\label{sec-ren}
To study one-loop ef\/fects in quantum f\/ield theory the background
f\/ield method is frequently used. One decomposes all f\/ields $\phi$
in their background values $\varphi$ and quantum f\/luctuations
$\delta\phi$, $\phi=\varphi +\delta\phi$. Then one expands the
classical action retaining the part which is quadratic in~$\delta\phi$ only. The result can usually be presented in the form
$\int d^mx\, (\delta \phi) D (\delta \phi)$ where the operator~$D$
depends on $\varphi$. A typical form of $D$ in NC theories is
given by (\ref{op1}). For example, in NC $\phi^4$ theory $l_1\sim
r_1 \sim \varphi \star \varphi$ and $l_2\sim r_2 \sim \varphi$.
The one loop ef\/fective action is given formally by $\ln \det (D)$,
and, in the framework of the zeta function regularization
\cite{Dowker:1975tf,Hawking:1976ja},
its' divergent part is proportional to an unsmeared heat kernel coef\/f\/icient
\begin{gather*}
W_{\rm div}=-\frac 1{2s} a_m(D),
\end{gather*}
where $s$ is a regularization parameter which should be taken $0$
at the end of the calculations. The divergences appearing in the
limit $s\to 0$ must be absorbed in redef\/initions of the parameters
in classical action. To make this possible, $a_m(D)$ should repeat
the structure of the classical action. (On a side note, let us
remark that in the framework of the spectral action principle~$a_m$, together with other coef\/f\/icients, {\it defines} the
classical action, cf.\ Section~\ref{sec-sac}). In the case of NC~$\phi^4$ on~$\mathbb{R}^4_\Theta$ planar heat kernel coef\/f\/icients
$a_4^L(D)$ and $a_4^R(D)$ satisfy this requirement, and the corresponding
mixed coef\/f\/icient vanishes, $a_4^{\rm mix}(D)=0$. We can conclude that
there are no problems with one-loop renormalization of NC $\phi^4$
on the plane
(see \cite{Vassilevich:2005vk} for details and further references).

On the NC torus $\mathbb{T}^4_\Theta$ the situation is more
subtle. By substituting the functional form of~$l_{1,2}$ and
$r_{1,2}$ given in the previous paragraph in (\ref{hkTm}) one
obtains (for pure Diophantine $\Theta$) non-local divergent terms
of the form $\int \varphi \cdot \int \varphi \star \varphi\star
\varphi$ and $(\int \varphi^2)^2$. By imposing anti-periodic
conditions on $\varphi$ one can achieve $\int\varphi=0$ and thus
kill one type of the divergences. Next, to achieve matching
between one-loop divergences and the classical action we add a
term $(\int \phi^2)^2$ to the classical action. Fortunately, this
procedure closes: the new term in the classical action does not
generate new types of the one-loop divergences. The
renormalizability at this order is restored~\cite{Gayral:2006vd}.

As an example of applications of the heat kernel expansion to
anomalies and anomalous actions in NC theory let us consider
calculations of an induced Chern--Simons action on
$\mathbb{T}^3_\Theta$ \cite{Vassilevich:2007gt}. The one-loop
ef\/fective action for 3-dimensional Dirac fermions is def\/ined
by determinant of the Dirac operator
\begin{gather*}
\slashed{D}=i\gamma^\mu \left( \partial_\mu +iL(A_\mu^L)+ i
R(A_\mu^R) \right). 
\end{gather*}
A peculiar feature of the approach we follow here is the presence
of two $U(1)$ gauge f\/ields, $A_\mu^L$~and~$A_\mu^R$, which are
naturally coupled to two conserved currents \cite{Vassilevich:2007gt}.
It is known
\cite{Alvarez-Gaume:1984nf,Niemi:1986kh,D1}
that the parity-violating part of the ef\/fective
action (the parity anomaly) reads
\begin{gather*}
W^{\rm pv}=i\frac {\pi}2 \eta (0),
\end{gather*}
where the $\eta$-function is def\/ined as a sum over the eigenvalues
$\lambda_n$ of the Dirac operator: $\eta (s)=\sum_{\lambda_n>0}
(\lambda_n)^{-s} -\sum_{\lambda_n<0}(-\lambda_n)^{-s}$. By repeating
the arguments known from the commutative case~\cite{Alvarez-Gaume:1984nf},
one can show that variation of the eta invariant under small variations
of the gauge f\/ields is
\begin{gather*}
\delta \eta (0)=-\frac 2{\sqrt{\pi}} a_2(\delta \slashed{D},\slashed{D}^2).
\end{gather*}
In the zeta regularization, $W^{\rm pv}=\frac 12 S_{\rm CS}$, where
$S_{\rm CS}$ is the Chern--Simons action. Now, let us suppose that
$\Theta^{12}=-\Theta^{21}\equiv \theta$, $\Theta^{13}=\Theta^{23}
=0$, and that $\theta /(2\pi)=P/Q$ is rational. Then, by combining
the formulae above with known heat kernel coef\/f\/icients on NC torus
(\ref{hkTm}) one obtains, that the Chern--Simons action in this case
consists of three terms $S_{\rm CS}=S_{\rm CS}^L+S_{\rm CS}^R
+ S_{\rm CS}^{\rm mix}$, where f\/irst two terms depend exclusively on
$A^L$ or $A^R$ respectively and coincide with induced Chern--Simons
actions on the NC plane \cite{Chu:2000bz,Grandi:2000av}. The third,
mixed, term reads
\begin{gather*}
S_{\rm CS}^{\rm mix}=-\frac i{2\pi} \int d^3x \, \sqrt{g}
\epsilon^{\mu\nu\rho}  [A_\mu^L]_Q\partial_\nu
[A_\rho^R]_Q.
\end{gather*}
It depends on the parts $[A_\mu^{L,R}]_Q$ of the gauge f\/ields
which are periodic in $x^1$ and $x^2$ with the period $(2\pi)/Q$.
Such f\/ields remind us of solid state physics and correspond to a
crystal consisting of $Q\times Q$ fundamental domains.

More examples of applications of the heat kernel expansion to
renormalization and anomalies in NC theories, as well as comparison
to the results obtained by other methods and further references
can be found in
\cite{Gayral:2004cs,Gayral:2004cu,Gayral:2005ih,Gayral:2006vd,
Grosse:2006hh,Strel:2006,Strelchenko:2007xh,Vassilevich:2004ym,
Vassilevich:2005vk,Vassilevich:2007gt}.

\section{Conclusions and open problems}
As we have already seen, the heat kernel expansion on NC spaces
looks much more complicated than in the commutative case. The very
structure of the expansion depends crucially on compactness of the
manifold and on whether $\Theta^{\mu\nu}$ is degenerate or not. In
the compact case, the number theory aspects play an important
role. Nevertheless, the case of the Moyal product and of f\/lat
noncommutative dimensions is understood quite well, though general
analytic formulae are not always available.

At the same time, very little is known about the case when the NC
coordinates are curved (i.e., when the Riemann metric depends on
these coordinates). Even a proper generalization of the Laplace
operator is still missing. The same is true for a generic position
dependent NC parameter $\Theta^{\mu\nu}$ which leads to a star
product of the Kontsevich type \cite{Kontsevich:1997vb}. The
problem seems to be deeper than just a technical one. In these two
cases, even classical actions, symmetries and the structure of
invariants, quantization rules are still to be determined. Note,
that understanding of the heat kernel expansion may in turn help
to solve the problems mentioned above (e.g., through the spectral
action principle).

\subsection*{Acknowledgements}
This work was supported in part by FAPESP (Brazil).

\pdfbookmark[1]{References}{ref}

\LastPageEnding


\begin{thebibliography}{99}

\footnotesize\itemsep=0pt

\bibitem{Alvarez-Gaume:1984nf}
  Alvarez-Gaume L., Della Pietra S., Moore G.W.,
  Anomalies and odd dimensions,
  {\it Annals Phys.}  {\bf 163} (1985), 288--317.

\bibitem{Avramidi:2005rn}
  Avramidi I.G.,
  Dirac operator in matrix geometry,
  {\it Int. J. Geom. Methods Mod. Phys.}  {\bf 2} (2005), 227--264,
\href{http://arxiv.org/abs/math-ph/0502001}{math-ph/0502001}.


\bibitem{Avramidi:2001tx}
  Avramidi I., Branson T.,
  Heat kernel asymptotics of operators with nonLaplace principal part,
  {\it Rev.\ Math.\ Phys.}\  {\bf 13} (2001), 847--890, \href{http://arxiv.org/abs/math-ph/9905001}{math-ph/9905001}.

\bibitem{Barvinsky:1990up}
  Barvinsky A.O., Vilkovisky G.A.,
  Covariant perturbation theory. 2.~Second order in the curvature. General
  algorithms,
  {\it Nuclear Phys.~B} {\bf 333} (1990), 471--511.


\bibitem{Branson:1990xp}
  Branson T.P., Gilkey P.B.,
  The asymptotics of the Laplacian on a manifold with boundary,
  {\it Comm. Partial Differential Equations}  {\bf 15} (1990), 245--272.

\bibitem{Branson:1995cm}
  Branson T.P., Gilkey P.B., Vassilevich D.V.,
  The asymptotics of the Laplacian on a manifold with boun\-dary.~2,
  {\it Boll.\ Union.\ Mat.\ Ital.}\  {\bf 11B} (1997), 39--67,
  \href{http://arxiv.org/abs/hep-th/9504029}{hep-th/9504029}.


\bibitem{Branson:1999jz}
  Branson T.P., Gilkey P.B., Kirsten K., Vassilevich D.V.,
  Heat kernel asymptotics with mixed boundary conditions,
  {\it Nuclear Phys.~B} {\bf 563} (1999), 603--626,
  \href{http://arxiv.org/abs/hep-th/9906144}{hep-th/9906144}.

\bibitem{Chamseddine:1996zu}
  Chamseddine A.H., Connes A.,
  The spectral action principle,
  {\it Comm.\ Math.\ Phys.}  {\bf 186} (1997), 731--750,
  \href{http://arxiv.org/abs/hep-th/9606001}{hep-th/9606001}.

\bibitem{Chamseddine:2007ia}
  Chamseddine A.H., Connes A.,
  A dress for SM the beggar,
 \href{http://arxiv.org/abs/0706.3690}{arXiv:0706.3690}.


\bibitem{Chu:2000bz}
  Chu C.S.,
  Induced Chern--Simons and WZW action in noncommutative spacetime,
  {\it Nuclear Phys.~B} {\bf 580} (2000), 352--362,
\href{http://arxiv.org/abs/hep-th/0003007}{hep-th/0003007}.


\bibitem{ConnesTm}
Connes A., $C^*$-alg\`ebres et g\'eom\'etrie dif\/f\'erentielle,
{\it C.R. Acad. Sci. Paris} {\bf 290} (1980), 599--604.

\bibitem{ConnesBook}
Connes A., Noncommutative geometry, Academic Press, London and San
Diego, 1994.

\bibitem{CL}
Connes A., Landi G., Noncommutative manifolds, the instanton
algebra and isospectral deformations, {\it Comm.\ Math.\ Phys.}
{\bf 221} (2001), 141--159, \href{http://arxiv.org/abs/math.QA/0011194}{math.QA/0011194}.

\bibitem{CDB}
Connes A., Dubois-Violette M., Noncommutative
f\/inite-dimensional manifolds. I. Spherical manifolds and related
examples, {\it Comm.\ Math.\ Phys.} {\bf 230} (2002), 539--579, \href{http://arxiv.org/abs/math.QA/0107070}{math.QA/0107070}.


\bibitem{D1}
  Deser S., Griguolo L., Seminara D.,
  Gauge invariance, f\/inite temperature and parity anomaly in $D = 3$,
  {\it Phys.\ Rev.\ Lett.}  {\bf 79} (1997), 1976--1979,
\href{http://arxiv.org/abs/hep-th/9705052}{hep-th/9705052}.


\bibitem{Dowker:1975tf}
  Dowker J.S., Critchley R.,
  Ef\/fective Lagrangian and energy momentum tensor in de Sitter space,
  {\it Phys.\ Rev.~D} {\bf 13} (1976), 3224--3232.

\bibitem{Essouabri:2007wi}
  Essouabri D., Iochum B., Levy C., Sitarz A.,
  Spectral action on noncommutative torus,
\href{http://arxiv.org/abs/0704.0564}{arXiv:0704.0564}.

\bibitem{Gayral:2004cs}
  Gayral V.,
  Heat-kernel approach to UV/IR mixing on isospectral deformation
  manifolds,
  {\it Ann. Henri Poincar\'e} {\bf 6} (2005), 991--1023,
  \href{http://arxiv.org/abs/hep-th/0412233}{hep-th/0412233}.


\bibitem{Gayral:2004cu}
  Gayral V., Gracia-Bondia J.M., Ruiz F.R.,
  Trouble with space-like noncommutative f\/ield theory,
  {\it Phys.\ Lett.~B}  {\bf 610} (2005), 141--146,
  \href{http://arxiv.org/abs/hep-th/0412235}{hep-th/0412235}.


\bibitem{Gayral:2005ih}
  Gayral V., Gracia-Bondia J.M., Ruiz F.R.,
  Position-dependent noncommutative products: classical construction and
  f\/ield theory,
  {\it Nuclear Phys.~B} {\bf 727} (2005), 513--536,
  \href{http://arxiv.org/abs/hep-th/0504022}{hep-th/0504022}.


\bibitem{GI:2004}
Gayral V., Iochum B.,
The spectral action for Moyal plane,
{\it J.\ Math.\ Phys.} {\bf 46} (2005), 043503, 17 pages, \href{http://arxiv.org/abs/hep-th/0402147}{hep-th/0402147}.

\bibitem{Gayral:2005ad}
  Gayral V., Iochum B., Varilly J.C.,
  Dixmier traces on noncompact isospectral deformations,
  {\it J.\ Funct.\ Anal.}  {\bf 237} (2006), 507--539,
  \href{http://arxiv.org/abs/hep-th/0507206}{hep-th/0507206}.

\bibitem{Gayral:2006vd}
  Gayral V., Iochum B., Vassilevich D.V.,
  Heat kernel and number theory on NC-torus,
  {\it Comm.\ Math.\ Phys.}  {\bf 273} (2007), 415--443,
  \href{http://arxiv.org/abs/hep-th/0607078}{hep-th/0607078}.

\bibitem{Gold}
Gilkey P.B., Invariance theory, the heat equation, and the
Atiyah--Singer index theorem, CRC Press, Boca Raton, FL, 1995.

\bibitem{Gbook}
Gilkey P.B., Asymptotic formulae in spectral geometry, Chapman \&
Hall/CRC, Boca Raton, FL, 2004.

\bibitem{Grandi:2000av}
  Grandi N.E., Silva G.A.,
  Chern--Simons action in noncommutative space,
  {\it Phys.\ Lett.~B} {\bf 507} (2001), 345--350,
  \href{http://arxiv.org/abs/hep-th/0010113}{hep-th/0010113}.

\bibitem{Grosse:2006hh}
  Grosse H., Wohlgenannt M.,
  Noncommutative QFT and renormalization,
  {\it J.\ Phys.\ Conf.\ Ser.}  {\bf 53} (2006), 764--792,
  \href{http://arxiv.org/abs/hep-th/0607208}{hep-th/0607208}.


\bibitem{Grosse:2007dm}
  Grosse H., Wohlgenannt M.,
  Induced gauge theory on a noncommutative space,
  \href{http://arxiv.org/abs/hep-th/0703169}{hep-th/0703169}.

\bibitem{Grosse:2004yu}
  Grosse H., Wulkenhaar R.,
  Renormalisation of $\phi^4$ theory on noncommutative ${\mathbb R}^4$ in the matrix
  base,
  {\it Comm.\ Math.\ Phys.}  {\bf 256} (2005), 305--374,
  \href{http://arxiv.org/abs/hep-th/0401128}{hep-th/0401128}.


\bibitem{Hawking:1976ja}
  Hawking S.W.,
  Zeta function regularization of path integrals in curved space-time,
  {\it Comm.\ Math.\ Phys.}  {\bf 55} (1977), 133--148.

\bibitem{Klausa5}
Kirsten K., The $a_5$ heat kernel coef\/f\/icient on a manifold with
boundary, {\it Classical Quantum Gravity} {\bf 15} (1998), L5--L12,
\href{http://arxiv.org/abs/hep-th/9708081}{hep-th/9708081}.

\bibitem{Kbook}
Kirsten, K., Spectral functions in mathematics and physics,
Chapman \& Hall/CRC, Boca Raton, FL, 2001.

\bibitem{Kontsevich:1997vb}
  Kontsevich M.,
  Deformation quantization of Poisson manifolds, I,
  {\it Lett.\ Math.\ Phys.}  {\bf 66} (2003), 157--216,
  \href{http://arxiv.org/abs/q-alg/9709040}{q-alg/9709040}.

\bibitem{Madore:1991bw}
  Madore J.,
  The fuzzy sphere,
  {\it Classical Quantum Gravity}  {\bf 9} (1992), 69--88.

\bibitem{Nest:2002ba}
  Nest R., Vogt E., Werner W.,
  Spectral action and the Connes--Chamsedinne model,
  {\it Lect.\ Notes Phys.}  {\bf 596} (2002), 109--132.


\bibitem{Niemi:1986kh}
  Niemi A.J.,
  Topological terms induced by f\/inite temperature and density f\/luctuations,
  {\it Phys.\ Rev.\ Lett.}  {\bf 57} (1986), 1102--1105.


\bibitem{Sasakura:2004dq}
  Sasakura N.,
  Heat kernel coef\/f\/icients for compact fuzzy spaces,
  {\it JHEP} {\bf 2004} (2004), no.~12, 009, 9~pages,
  \href{http://arxiv.org/abs/hep-th/0411029}{hep-th/0411029}.

\bibitem{Sasakura:2005px}
  Sasakura, N.,
  Ef\/fective local geometric quantities in fuzzy spaces from heat kernel
  expansions,
  {\it JHEP} {\bf 2005} (2005), no.~3, 015, 26 pages,  \href{http://arxiv.org/abs/hep-th/0502129}{hep-th/0502129}.

\bibitem{Strel:2006}
Strelchenko A.,
Heat kernel of non-minimal gauge f\/ield kinetic operators on Moyal plane,
  {\it Internat. J. Modern Phys.~A}  {\bf 22} (2007), 181--202,
  \href{http://arxiv.org/abs/hep-th/0608134}{hep-th/0608134}.

\bibitem{Strelchenko:2007xh}
  Strelchenko A.V., Vassilevich D.V.,
  On space-time noncommutative theories at f\/inite temperature,
{\it Phys. Rev.~D} {\bf 76} (2007), 065014, 12 pages,
  \href{http://arxiv.org/abs/0705.4294}{arXiv:0705.4294}.

\bibitem{Szabo:2006wx}
  Szabo R.J.,
  Symmetry, gravity and noncommutativity,
  {\it Classical Quantum Gravity}  {\bf 23} (2006), R199--R242,
  \href{http://arxiv.org/abs/hep-th/0606233}{hep-th/0606233}.

\bibitem{Vassilevich:1994we}
  Vassilevich D.V.,
  Vector f\/ields on a disk with mixed boundary conditions,
  {\it J.\ Math.\ Phys.}  {\bf 36} (1995), 3174--3182,
  \href{http://arxiv.org/abs/gr-qc/9404052}{gr-qc/9404052}.


\bibitem{Vassilevich:2003xt}
  Vassilevich D.V.,
  Heat kernel expansion: user's manual,
  {\it Phys.\ Rep.}  {\bf 388} (2003), 279--360,
  \href{http://arxiv.org/abs/hep-th/0306138}{hep-th/0306138}.

\bibitem{Vassilevich:2003yz}
  Vassilevich D.V.,
  Non-commutative heat kernel,
  {\it Lett.\ Math.\ Phys.}\  {\bf 67} (2004), 185--194,
  \href{http://arxiv.org/abs/hep-th/0310144}{hep-th/0310144}.

\bibitem{Vassilevich:2004id}
  Vassilevich D.V.,
  Spectral problems from quantum f\/ield theory,
  {\it Contemp.\ Math.}  {\bf 366} (2005), 3--22,
 \mbox{\href{http://arxiv.org/abs/math-ph/0403052}{math-ph/0403052}}.

\bibitem{Vassilevich:2004ym}
  Vassilevich D.V.,
  Quantum noncommutative gravity in two dimensions,
  {\it Nuclear Phys.~B}  {\bf 715} (2005), 695--712,
  \href{http://arxiv.org/abs/hep-th/0406163}{hep-th/0406163}.

\bibitem{Vassilevich:2005vk}
  Vassilevich D.V.,
  Heat kernel, ef\/fective action and anomalies in
  noncommutative theories,
  {\it JHEP} {\bf 2005} (2005), no.~8, 085, 19 pages,
  \href{http://arxiv.org/abs/hep-th/0507123}{hep-th/0507123}.

\bibitem{Vassilevich:2007gt}
  Vassilevich D.V.,
  Induced Chern--Simons action on noncommutative torus,
  {\it Modern Phys.\ Lett.~A}  {\bf 22} (2007), 1255--1263,
  \href{http://arxiv.org/abs/hep-th/0701017}{hep-th/0701017}.


\end{thebibliography}
\end{document}